\documentclass{article}
\usepackage{spconf,amsmath,graphicx}
\usepackage{tipa}
\usepackage{etoolbox}
\usepackage[utf8]{inputenc}
\usepackage[english]{babel}
\usepackage{adjustbox}
\usepackage{amssymb}
\usepackage{eqnarray,amsmath}
\usepackage{amsmath}
\usepackage{amsthm}
\usepackage{mathtools}
\usepackage{relsize}
\usepackage{ellipsis}
\usepackage{textcomp}
\usepackage{cite}
\usepackage{graphicx}  
\usepackage{subfigure}
\usepackage{epsfig}
\usepackage{epstopdf}
\usepackage{booktabs,tabularx}
\usepackage{caption}
\usepackage{psfrag}
\usepackage{grffile}

\usepackage{color}
\usepackage{hyperref}
\usepackage[utf8]{inputenc}
\usepackage[english]{babel}
\usepackage{amsthm}
\usepackage{makecell}

\usepackage{algpseudocode}
\usepackage{algorithm}

\title{Svadhyaya System for the Second Diagnosing COVID-19 Using Acoustics Challenge 2021}
\name{\parbox{\linewidth}{\centering
Deepak Mittal$^1$, Amir H. Poorjam$^1$, Debottam Dutta$^2$, Debarpan Bhattacharya$^2$, Zemin Yu$^1$, \\ Sriram Ganapathy$^2$, Maneesh Singh$^1$ }}

\address{$^1$Verisk Analytics Inc., Jersey City, NJ, US \\
$^2$LEAP Lab, Indian Institute of Science, Bangalore, India \\
\parbox{\linewidth}{\centering
\small\texttt{Email:\{deepak.mittal, amir.poorjam, zemin.yu, maneesh.singh\}@verisk.com} \\ 
\small\texttt{\{debottamd, debarpanb, sriramg\}@iisc.ac.in}}
}

\begin{document}

%
%
%

%
\maketitle

\begin{abstract}
    This report describes the system used for detecting COVID-19 positives using three different acoustic modalities, namely speech, breathing, and cough in the second DiCOVA challenge. The proposed system is based on the combination of $4$ different approaches, each focusing more on one aspect of the problem, and reaches the blind test AUCs of $86.41$, $77.60$, and $84.55$, in the breathing, cough, and speech tracks, respectively, and the AUC of $85.37$ in the fusion of these three tracks.
\end{abstract}
\begin{keywords}
COVID-19, acoustics, filter-bank learning, respiratory diagnosis, healthcare
\end{keywords}
\section{System Description}
\label{sec:intro}
The proposed system used in each track of the challenge is an ensemble of four approaches, each focusing on one aspect of the problem. The first approach is based on the bidirectional LSTM (BLSTM) classifier. The second approach uses the BLSTM in the first approach but focuses on improving the generalization of the model by using a linear combination of the training data. In the third approach, we replace the filter-bank used in the front-end of the BLSTM with a learnable filter-bank and a relevance weighting mechanism to provide a better signal representation for the acoustic signals. In the final approach, we use the Time Delay Neural Network (TDNN) \cite{snyder2018x} and the Emphasized Channel Attention, Propagation and Aggregation in TDNN (ECAPA-TDNN) \cite{desplanques2020ecapa} model either as a feature extractor to map an audio signal into a fixed-length vector, or as a classifier to detect COVID-19 positive samples. These approaches are then linearly combined at the probability score level.

\subsection{BLSTM Classifier}
This approach is based on extraction of log-mel spectogram from the breathing, cough and speech sounds and feeding them into a BLSTM classifier for COVID classification. This system is provided as the baseline system by the organisers\cite{dicova2_baseline}. 
\subsection{Mixup Training}
One important problem in deep learning models is that when the distribution of the test samples changes, even slightly, from that of the training data, the network is not able to extrapolate because of the overfitting on training data distribution. Zhang et al. \cite{zhang2018mixup} addressed the issue by proposing the mixup technique for augmentation and to improve the generalization capability of the neural network. They have shown that the performance improves for a set of diverse datasets in a variety of applications such as image and speech. The mixup is applied to the BLSTM to improve the generalisation of the model. In this technique, the training samples and the corresponding labels are linearly combined to form new training samples and target labels. This helps the model to learn the distributions which are linearly related to the training data distribution. Therefore, the model performs better for test set, whose distribution is different than training set, but linearly related to it.

\subsubsection{Algorithm}
The mixup of the training samples are done as the following:
\begin{equation}
x^{\prime} = \lambda x_m + (1-\lambda) x_n
\end{equation}
\begin{equation}
y^{\prime} = \lambda y_m + (1-\lambda) y_n
\end{equation}
where $\lambda$ is the coefficient for convex combination, $x_m$, $x_n$ are different training samples and $y_m$, $y_n$ are the corresponding target labels. We notice that the target labels become fractional when one positive and one negative class are mixed up. Although the task does not remain a binary classification task, the binary cross-entropy loss function can be used for computation of the loss corresponding to the fractional labels as follows:
\begin{equation}
    \mathcal{L}_{\mathrm{mixup}}(\hat{y}, y^\prime) = \lambda.\mathcal{L}(\hat{y}, y_m) + (1-\lambda).\mathcal{L}(\hat{y}, y_n)
\end{equation}
where $\hat{y}$ is the prediction probability, $y_m$ and $y_n$ are binary targets. The $\lambda$ for mixup is picked from the $Beta(\alpha, \alpha)$ distribution. The $\alpha$ is a hyperparameter for the model.
The algorithm for the implementation of mixup training is summarized in algorithm \ref{algo_mixup}.
\begin{algorithm}[h]
	\caption{Mixup Training Algorithm}
    \label{mixup}
	\begin{algorithmic}[1]
		\State $X$ $\rightarrow$ data matrix input in a mini batch
		\Comment {\%size: $m\times n$\%}
		\State Y $\rightarrow$ label vector input in a mini batch
		\Comment {\%size: $m\times1$\%}
		\State $\alpha$  $\rightarrow$ $Beta(\alpha, \beta)$ distribution parameter with $\beta=\alpha$\\
		\For{minibatch}
		\State $\lambda \sim Beta(\alpha, \alpha)$ \Comment {\%size: $m\times1$\%}
		\State $X^\prime \rightarrow \lambda .X + (1-\lambda).R(X)$\\
		\Comment {\%size: $m\times n$\%}
		\State $Y^\prime \rightarrow \lambda .Y + (1-\lambda).R(Y)$\\
		\Comment {\%size: $m\times1$\%}
		\State $\hat{Y}$ $\rightarrow$ $Model(X^\prime)$\Comment {\%forward pass\%}
		\State loss = $\mathcal{L}_{\mathrm{mixup}}(\hat{Y}, Y^\prime)$\Comment {\%as in eqn. 3\%}
		\State loss.backward()
		\State optimizer.step()
		\EndFor
	\end{algorithmic}
	\label{algo_mixup}
\end{algorithm}\\
In algorithm \ref{algo_mixup}, the $R(X)$ randomly shuffles the rows of the matrix X. 
\subsubsection{Softmax With Temperature}
The temperature can be applied on the final softmax activation to increase the model performance as pointed out by Berthelot et al. \cite{berthelotmixmatch}.
Given that $\{z_1,\ldots,z_i,\ldots,z_n\}$ denotes the unscaled output of the final layer of the network, having $n$ output nodes, the softmax with temperature output at $i^{\mathrm{th}}$ output node is represented as
\begin{equation}
    y_i = \frac{\exp{(\frac{z_i}{\tau})}}{\sum_{j=1}^{n}\exp{(\frac{z_j}{\tau})}}.
\end{equation}
Lower value of $\tau$ boosts the ``confidence'' of the model towards it's predictions. This is also observed during the experimentation with different values of $\tau$. When temperature, in addition with mixup, is applied to the baseline BLSTM model\cite{dicova2_baseline}, a reasonable improvement compared to the baseline performance is observed.

\subsubsection{Hyperparameter Tuning}
Table~\ref{tab: mixup_hyper} shows the impact of different $\alpha$ and $\tau$ values on the performance of the model in terms of the validation AUC. For the classifier, the hyper-parameters are kept the same as those used in the baseline system.
\begin{table}[th]
    \centering
    \caption{The impact of different temperature values on the performance of the model. The results are in terms of the AUC averaged over 5 validation folds. The numbers in the parenthesis are the baseline AUCs.}
    \label{tab: mixup_hyper}
    \begin{tabular}{|l|c|c|c|}
    \hline
     Modality & $\alpha$ & $\tau$ & Average Val AUC\\ \hline
     Breathing &0.4 &  0.001 & 77.80 (77.42)\\
    \hline
     Cough &0.4& 0.1 & 77.00 (75.64)\\
    \hline
    Speech &0.4 & 0.01 & 81.24 (79.6)\\
    \hline
    \end{tabular}
\end{table}

\subsection{Filter-bank Learning and Relevance Weighting}
In this approach we consider a learnable front-end trained directly on the raw waveforms to learn task-specific representations. The learnable front-end incorporates two main parts, a learnable Gaussian filter-bank layer and a relevance weighting network. 
\subsubsection{Acoustic Filter-bank Layer}
Input to this layer are $T$ frames, each with $S$ raw audio samples with a matrix of dimension $S\times T$. This matrix of raw samples is passed through a 1D conv layer whose kernels are parameterized as cosine modulated Gaussian functions, as described in \cite{agrawal2020interpretable}, 
\begin{equation}
    g_i(l) = \cos{2\pi \mu_i l} \times \exp{(-l^2\mu_i^2/2)}, 
\end{equation} 
where $l$ is the discrete time index, $g_i(l)$ is the i-th kernel $(i = 1,2,..,F)$ and $\mu_i$ is the centre frequency of the $i^{\mathrm{th}}$ kernel. The kernels perform convolution in each frame and generate $F$ feature maps. These feature maps are then squared, average pooled and log transformed. This way, from each frame, an $F$ dimensional feature is obtained. Stacking these feature vectors from $T$ frames, an $F\times T$ representation $\boldsymbol{x}$ is obtained which we call the learned time--frequency representation. 
\subsubsection{Relevance Weighting}
For relevance weighting, we consider a sub-network approach as described in \cite{dutta2021multihead}. Input to this layer is the learned representation $\boldsymbol{x}$. The relevance network is a $2$ layer fully connected network with sigmoidal activation in the hidden and output layers. The relevance network receives one element from each time-frequency bin along with a temporal context of $(2c +1)$ from that sub-band to generate the relevance weight for that time-frequency element \cite{dutta2021multihead}. 
\begin{equation}
    W_{k,j} = \sigma(\boldsymbol{\Omega}_{2}(\sigma(\boldsymbol{\Omega}_{1}\boldsymbol{y}_{k,j}+\boldsymbol{b}_{1}) + b_{2}))
\end{equation}
where $ W_{k,j}$ is the entry in $k^{th}$ row and $j^{th}$ column of the weight mask $\boldsymbol{W}$, $\boldsymbol{y}_{k,j}$ is the $2c+1$ dimensional input vector $[x_{k,(j-c)}, ..,x_{k,j}, .., x_{k,(j+c)}]^T $, $\boldsymbol{\Omega}_1,\boldsymbol{b}_1$ and $\boldsymbol{\Omega}_2,b_2$ are weight matrices and biases for the first and second layer respectively. The same network is shared across all t-f bins to generate the weight mask $\boldsymbol{W}$. The mask, $\boldsymbol{W}$ is multiplied element-wise with $\boldsymbol{x}$ to obtain a enhanced t-f representation $\boldsymbol{x}_W$ which is then fed as an input to a classifier. This learnable front-end and the classifier is trained jointly for the downstream task with a supervised loss. 
The classifier used for this model is a BLSTM followed by a FC network with $64$ units in the hidden layer \cite{dicova2_baseline}. 

\subsubsection{Hyperparameter Tuning}
\begin{figure}[t!]
        \centering
	    \includegraphics[width=0.5\textwidth]{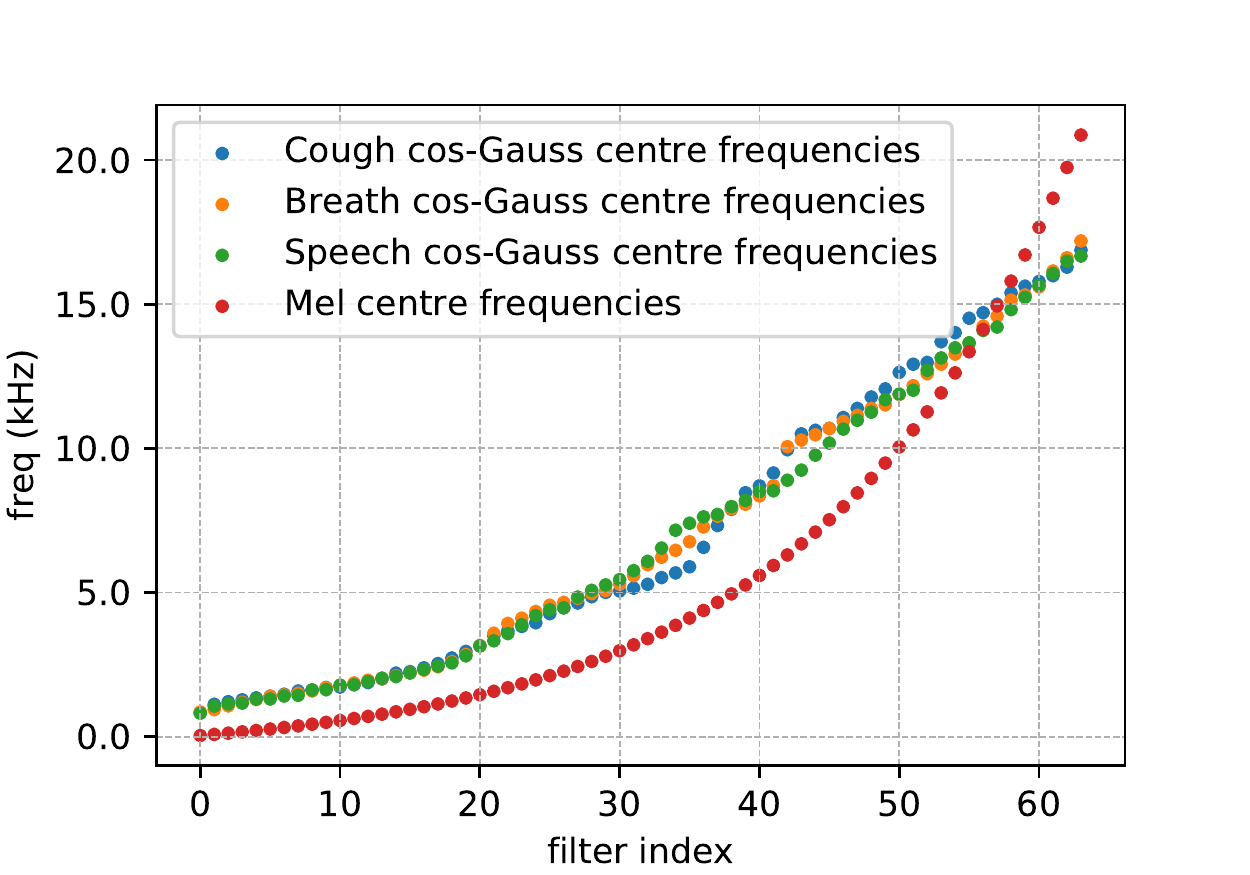}
	    \vspace{-0.2cm}
	    \caption{Distribution of the centre frequencies of mel filterbank (red) and learnt Cos-Gauss centre frequencies for  cough (blue), breathing (orange) and speech (green).}
	    \label{fig:center_freq}
\end{figure}

For this model, we consider the number of Gaussian filters, $F = 64$ and the kernel size $k = 353$ ($8$ ms  for  audio  signal  sampled  at $44.1$ kHz). The relevance network is a simple FC network with hidden layer size $50$. In our experiments, we remove the skip-connection and multi-head relevance described in \cite{dutta2021multihead}, as the model showed better performance in a single head setting without skip-connection for this dataset. The context parameter, $c=51$ is used after tuning on the development set. The window length, $S$ and frame length, $T$ are kept at $1102$ (with hop length of $441$ samples) and $51$ to match with the DiCOVA-2 baseline BLSTM model. Fig~\ref{fig:center_freq} shows the learned centre frequency distributions from all the sound categories. We can observe that the learned filter-banks put more filters in the middle frequency range compared to the mel filter-bank. Also the learned filter-banks show a more linear distribution of filters and omits very high frequencies. 
Table \ref{table:relev_res} shows the performance of the model on different sound categories and compares them with the corresponding baseline model. We observe that the mean validation AUC improves across all the modalities for the relevance weighted representations (Cos-Gauss + relev.) from the learned front-end. 

\begin{table}[th]
\caption{Performance comparison of the two models with relevance weighted Cos-Gauss representations and mel representations in terms of AUC averaged over the five folds of the development data set.}
\centering
\label{table:relev_res}
\begin{tabular}{|c|c|c|}
\hline
 Modality & Cos-Gauss + relev. & Mel (Baseline)\\
\hline
Breathing & 78.43 & 77.42 \\ 
\hline
Cough & 77.42 & 75.64  \\
\hline 
Speech & 81.52 & 79.60 \\
\hline

\end{tabular}
\end{table}

\subsection{TDNN}
\label{ssec: tdnn}
In this approach, we explore the Time Delay Neural Network (TDNN) architecture \cite{snyder2018x}, as well as the Emphasized Channel Attention, Propagation and Aggregation in TDNN (ECAPA-TDNN) architecture \cite{desplanques2020ecapa}. The latter is an extended version of the original x-vector system proposed to extract speaker representations for the speaker verification problem. Therefore, we not only consider the model as an extractor for representations but also train the architecture as a classifier using the training set. In other words, as an extractor, we directly use the embeddings as our features; as a classifier, we train a representation more specific to the task of detecting COVID-19 and make predictions accordingly. 

\subsubsection{Extractor}
\label{sssec: extractor}
The motivation for this method is to apply transfer learning based on some existing pre-trained audio classification models, with the goal of experimenting whether the extracted features contain useful information in order to perform an entirely different classification task. We consider one of the published systems by SpeechBrain \cite{speechbrain} on the audio classification task: Language Identification from Speech Recordings with ECAPA embeddings on CommonLanguage. We use the system to generate audio embeddings, and then apply various machine learning models for detecting COVID-19 positive samples. 

The embedding acquired from the system above has a vector length of 192. Then we standardize features by removing the mean and re-scaling to unit variance. We consider a dimension reduction step afterwards by applying principal component analysis and tune the number of components to keep ($n$). We also look into both logistic regression and support vector machine as our models for the classification task. For the logistic regression, we use the L2 penalty term and tune the regularization parameter ($C$) and the weights associated with classes (weight). For the support vector machine, we tune the kernel type (kernel), the regularization parameter ($C$) and the weights associated with classes (weight). Finally, We perform randomized search on all the parameters described. 

Support vector machine has the highest performance on the breathing and cough modalities, while logistic regression reaches the best achievement on the speech modality. Table \ref{table:extractor} shows the detailed parameters used for the best performing model for each modality. We can learn that most of features are kept during the dimension reduction step and better results seem to be achieved by stronger regularization. 

\begin{table}[th]
\caption{Parameters for the best performing model for each modality. $C$ is in log scale. AUC is averaged over 5 validation folds. }
\centering
\label{table:extractor}
\begin{tabular}{|c|c|c|c|c|c|c|}
\hline
Modality & $C$ & kernel & weight & $n$ & AUC \\
\hline
Breathing & 0 & RBF & None & 184 & 77.45 \\ 
\hline
Cough & -2.75 & Linear & Balanced & 190 & 73.68  \\
\hline 
Speech & -1.5 & N/A & None & 181 & 76.65  \\
\hline
\end{tabular}
\end{table}

\subsubsection{Classifier}
\label{sssec:classifier}
The extractor method described in Section \ref{sssec: extractor} provides us with reasonably valid performance. Therefore, we would like to take an additional step to train the entire classifier from scratch in the hope that a more specified representation would yield better validation results. In this approach, we consider the original TDNN architecture as our model. It is trained on the speech modality only, using the 5-fold training and validation sets provided by the organizers. 

The experiment is set up using a similar fashion as the original architecture but only to predict COVID-19 positive or not instead of speaker identification. Cross entropy loss is used and the best model is selected using the lowest validation loss over the entire training progress. We end up keeping the same configuration that contains 5 frame-level layers, 1 statistics pooling layer, 2 segment-level layers and the softmax layer in the end. However, We reduce the segment-level layer dimensions to 128 before feeding into the softmax layer for classifications because we only have two output classes. Moreover, step annealing technology is used for the learning rate schedules decay. It behaves like an exponential decay, but decreases in steps. Each model is trained for 48 epochs where the learning rate starts at 0.001, reduces by a factor of 0.9085 for each 2 epochs, and decreases to $\frac{1}{10}$ of the starting value in 24 epochs. 

The system is able to reach 78.35 average validation AUC for the speech modality, compared to 76.65 in Section \ref{sssec: extractor}. This shows a sizable improvement without even taking advantage of all the enhancements made to the original TDNN architecture. It also allows us to understand that learning representations directly from the data is an important factor for the performance gain.  

\subsection{Model Combination}
\label{ssec: combination}
Each of the models described above focuses on one limitation of the baseline system. Therefore, they are expected to provide complementary information and, consequently, to improve the overall performance \cite{avila2021investigating}.
To this aim, for each acoustic modality, we linearly combined the models at the score level. Given the vector of scores of the $N$ number of test samples, obtained by applying the $h^{\mathrm{th}}$ model, the final probability score vector is calculated as
\begin{equation}
    \textbf{V}_{1:N} = \sum_{h=1}^{H} a_h \textbf{v}_h
\end{equation}
where $a_h$ are the weights where $\sum_{h=1}^Ha_h$ = $1$ and $a_h\sim\mathrm{Dir}(\gamma)$ are drawn from a symmetric Dirichlet distribution of the concentration parameter $\gamma$. Setting $\gamma=0.4$ and sampling from the Dirichlet distribution 500 times, we found the optimal mixing weights based on the average validation AUCs.
Prior to the combination, the scores are normalized using z-score normalization.

\section{Results}
We developed our models based on The Second DiCOVA Challenge \cite{dicova2_baseline} development data set. This data set comprises of $3$ sound modalities, namely breathing, cough, and speech with $965$ sound samples of which $17.2 \%$ belong to COVID-19 positive class. This extreme class imbalance reflects the typical real world COVID-19 scenario.
The three sound categories come with total durations of $4.62$, $1.68$ and $3.93$ hours for breathing, cough, and speech, respectively. For our experiments we used the same five-fold split provided by the organizers. Finally, all developed models are evaluated on the Challenge leaderboard with a blind test set of $471$ samples for each category. 
Table~\ref{tab: final_results} shows the comparison between the baseline and the proposed systems for three acoustic modalities and the fusion of them in terms of the AUC. 

\begin{table}[th]
    \centering
    \caption{Performance comparison between the baseline and the proposed systems for different modalities in terms of AUC.}
    \label{tab: final_results}
    \begin{tabular}{|l|c|c|}
    \hline
     Modality & Test AUC & Baseline AUC\\ \hline
     Breathing & 86.41 & 84.50 \\
    \hline
     Cough &77.60& 74.89\\
    \hline
    Speech &84.55 & 84.26 \\
    \hline
    Fusion & 85.37 & 84.70\\
    \hline
    \end{tabular}
\end{table}

\bibliographystyle{IEEEbib}
\bibliography{strings,refs}

\end{document}